\documentclass[runningheads]{llncs}
\usepackage{enumitem}

\usepackage{graphicx}
\usepackage{float}
\usepackage{color}
\usepackage{xcolor}
\usepackage{subcaption}
\definecolor{mydarkgreen}{RGB}{0, 100, 0}
\usepackage{booktabs}
\usepackage{caption}
\usepackage{multirow}
\usepackage{multicol}
\usepackage{lipsum}
\usepackage{mwe}
\usepackage{xspace}
\newcommand{\approach}{NEP\-TUNE$^+$\xspace}

\usepackage{comment}
\usepackage[font=footnotesize]{caption} 
\usepackage{algorithm}
\usepackage{algpseudocode}
\usepackage{tabularx}
\usepackage{array}
\usepackage{booktabs}
\usepackage{threeparttable}
\usepackage{comment}

\newcommand{\blue}[1]{\textcolor{black}{#1}}

\begin{document}

\title{Dependency-aware Resource Allocation \\for Serverless Functions at the Edge}
%
%
\author{
Luciano Baresi 
\and Giovanni Quattrocchi
\and Inacio Ticongolo 
}

\authorrunning{Baresi, Quattrocchi, Ticongolo}
\institute{Dipartimento di Elettronica, Informazione e Bioingegneria,
\\ Politecnico di Milano, Italy \\
\email{\{name.surname\}@polimi.it}}

%
%
\maketitle              

\newlist{questions}{enumerate}{2}
\setlist[questions,1]{label*=\textbf{RQ\arabic*},ref=RQ\arabic*}
\setlist[questions,2]{label=(\alph*),ref=\thequestionsi(\alph*)}

\begin{abstract}

Serverless computing allows developers to break their code into small components, known as functions, which are automatically managed by a service provider. Being lightweight and modular, serverless functions have been increasingly employed in edge computing, where quick responses and adaptability are key to meeting strict latency requirements. In particular, edge nodes are intrinsically resource-constrained, and efficient resource allocation strategies are crucial for optimizing their usage. Different approaches exist in the literature, but they often overlook the dependencies among functions, that is, how and when functions invoke other functions, obtaining suboptimal results. 

This paper presents \approach, a dependency-aware resource (CPU cores) allocation solution for serverless functions deployed at the edge. The approach extends NEPTUNE, an existing framework for managing edge infrastructures, with a new theoretical model and control algorithm that take dependencies into account function. We evaluated \approach by using three applications and it is able to allocate up to 42\% fewer cores compared to NEPTUNE.

\keywords{serverless \and edge computing \and function dependencies \and resource allocation}
\end{abstract}

\section{Introduction}
Complex applications are increasingly built as sets of independent components, such as microservices~\cite{bhasi2021Kraken} to foster agility and speed in both development and runtime management. This high degree of modularization allows for independent management, but complicates communication among components and affects their performance. \blue{For this reason, understanding the logical dependencies among components is essential for the efficient management of the system~\cite{esparrachiari2018tracking}.}

Serverless computing is arising as a new instance of such highly modularized architectural paradigms~\cite{vojdan2021iotServLR}. It promotes the creation of applications as a collection of ``small'' functions~\cite{li2022kneeScale}, which are usually executed in lightweight containers to ensure their fast and efficient management \cite{cassel2022serverlessLR}.
Serverless functions are designed to be independently developed, deployed, and scaled automatically by a service provider. This high degree of flexibility allows for fast re-configuration and scaling in response to changes in the system workload and contributes to overall system agility.

Given these characteristics, serverless functions have been increasingly employed in edge computing~\cite{vojdan2021iotServLR}. In this context, applications are often constrained by strict latency requirements.  The inherent agility and ability to rapidly scale individual components allow serverless platforms to quickly adapt to workload fluctuations, facilitate the prompt execution of functions, and allow the system to meet latency requirements more effectively~\cite{wang2021lass}. However, it is essential to acknowledge that serverless functions can depend on other functions, which can significantly impact their performance and management~\cite{tarek2018costless}.

In the last few years, serverless computing has been widely studied as means to improve the management of applications deployed on edge infrastructures~\cite{gadepalli2019challengsOpServEdge}. Some approaches tackle the intelligent placement of edge functions on resource-limited nodes~\cite{cassel2022serverlessLR,el2021platformsServReview}; others focus on optimizing resource allocation~\cite{duarte2018servFuncAllocationIoT}. Yet, these approaches often overlook function dependencies, a crucial factor for performance modeling~\cite{xu2023statefulServ}. 

This paper introduces \approach, a solution that focuses on resource allocation for serverless functions deployed at the edge. \approach extends an existing edge framework, called  NEPTUNE~\cite{baresi2022neptune}, which allows for the smart placement and allocation of serverless functions but does not consider function dependencies.
\approach extends the theoretical model of NEPTUNE by proposing i) a new formalization of the problem that encodes function dependencies as an annotated Direct Acyclic Graph (DAG), and ii) a novel control algorithm that exploits the function dependency graph to save resources.
A comprehensive empirical evaluation compared NEPTUNE and \approach by means of three benchmark applications. Obtained results show that \approach allocates up to $42\%$ fewer cores than NEPTUNE, with comparable performance in terms of response times.

The rest of the paper is organized as follows. Section~\ref{sec:neptune} introduces NEPTUNE and highlights its limitations. Section~\ref{sec:solution} describes our solution, along with the new problem formulation and control algorithm. Section~\ref{sec:evaluation} presents the evaluation and discusses the results. Section~\ref{sec:related} surveys the related work, and Section~\ref{sec:conclusions} concludes the paper.

\section{NEPTUNE in a nutshell}
\label{sec:neptune}
Serverless computing is the driver of a significant paradigm shift that frees developers from infrastructure management~\cite{li2022kneeScale}, and some  approaches~\cite{xu2023statefulServ,daw2020xanadu} explored this paradigm for deploying and managing applications in edge infrastructures. To the best of our knowledge, only NEPTUNE~\cite{baresi2022neptune} provides a comprehensive and holistic management approach that considers network partitioning, placement, request routing, and the combined dynamic allocation of memory, CPUs, and GPUs.

NEPTUNE requires the code of the function to deploy, a threshold (service level agreement or SLA) on its response time, and the identification of the memory required for proper execution. 
The management exploits a three-level hierarchy: topology, community, and node. The global network \textit{topology} is split into a set of independent \textit{communities}. Each community is composed of edge \textit{nodes} (or servers) that are close to each other, that is, their network inter-delays are smaller than a set threshold. Each community is managed by a dedicated controller that takes into account user mobility, workload provenance, \blue{and the memory requirements for each function}. This controller exploits an optimization problem based on Mixed-Integer Programming to calculate the best placement of function instances and a set of routing policies that minimize network latency. Function placement implies deciding how many function instances are needed for each used function and the best node to host each of them. Since each node cannot always host all the instances and handle all the incoming workload, NEPTUNE uses routing policies to compute the request fraction to be routed to other nodes. The same formulation is used to first handle the workload that can be accelerated through GPUs, and then, the remaining workload is assigned to CPUs exclusively.

\blue{Whereas GPUs and memory are entirely managed by the community controller}, the node level controller oversees the proper execution of requests \blue{on CPUs}, making sure that each function instance is provisioned with enough cores to complete executions within the given SLA.
A lightweight Proportional Integer (PI) controller is attached to each function instance $f_i$ with the goal of keeping the response time close to a given set point:

\begin{equation}
\label{eq:setpoint}
sp_i = \alpha * SLA_i
\end{equation}

\noindent
where $\alpha$ is a scaling parameter ($0<\alpha \le 1$). The more $\alpha$ is close to $1$, the more the response time is kept close to SLA, with a risk for potential violations. Conversely, if $\alpha$ is significantly lower than $1$, the controller ensures better performance, but more resources are needed. Therefore, $\alpha$ represents a tunable trade-off between performance and resource utilization.
\begin{figure}[t]
  \centering
\includegraphics[width=0.7\textwidth]{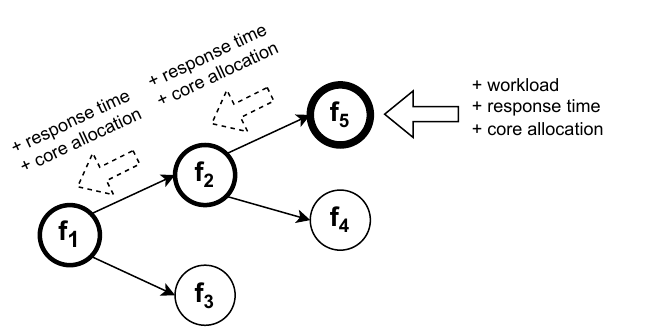}
  \caption{Example application with function dependencies.}
  \label{fig:example}
\end{figure}
\subsection{Limitations}
\label{sec:netpune:limitations}

NEPTUNE does not consider function dependencies, which can lead to inefficient resource allocation. Let us consider, for example, the application in Figure~\ref{fig:example}. This application consists of five functions, $f_1, f_2, f_3, f_4, f_5$, with their dependencies outlined in a directed acyclic graph (DAG): $f_1$ depends on $f_2$ and $f_3$, while $f_2$ depends on $f_4$ and $f_5$.

If we suppose that function $f_5$ is supposed to manage a workload spike, its response time suddenly increases, and its local node level controller is prompted to augment allocated cores. While $f_5$'s controller stabilizes the situation and brings the response time closer to the set point, the response times of $f_2$, and then of $f_1$, also grow, given their dependency on $f_5$.

The inefficiency of NEPTUNE lies in the behavior of the controllers associated with $f_2$ and $f_1$'s: higher response times, due to the slow responses from $f_5$, imply increasing the cores allocated to $f_1$ and $f_2$. This reaction is redundant, as the issue does not stem from either $f_2$ or $f_1$, but the bottleneck is $f_5$. Instead of allocating extra cores to $f_2$ and $f_1$, these resources would have been better utilized to speed up $f_5$. 

This is why NEPTUNE must be improved/extended to address this limitation and fix redundant allocations: \blue{CPU cores} must be allocated efficiently even in the presence of dependent serverless functions.

\section{Solution}
\label{sec:solution}
\approach extends NEPTUNE with a new theoretical model and control algorithm to efficiently allocate resources to serverless functions with dependencies. \blue{In particular, \approach aims to improve the allocation of CPUs cores at the node level in light of the limitations described in Section \ref{sec:netpune:limitations}, while it inherits from NEPTUNE its placement strategy that minimizes network delays and the management of GPUs and memory.}

\subsection{Theoretical Model}
\label{sec:functionDepModel}

Let $F$ be a set of serverless functions whose dependencies are encoded as a DAG where nodes are the function instances and edges are the invocations between functions. This DAG is assumed to be either manually defined by the user or automatically generated utilizing network or log analysis techniques \cite{ghirotti2018tracking}.

We do not consider invocation cycles (i.e., we employed DAGs for modeling dependencies), in line with the recommendations against their use, as suggested by Fontana et al.~\cite{Fontana2016ICSME}.
A cycle denotes a situation where a function indirectly relies on its own output to commence its execution, forming an untenable loop. This could lead to an endless cycle of executions or deadlock scenarios that are unmanageable cost-wise on real serverless platforms~\cite{taibi2018DAG}. 

\begin{figure}[t]
  \centering
\includegraphics[width=0.55\textwidth]{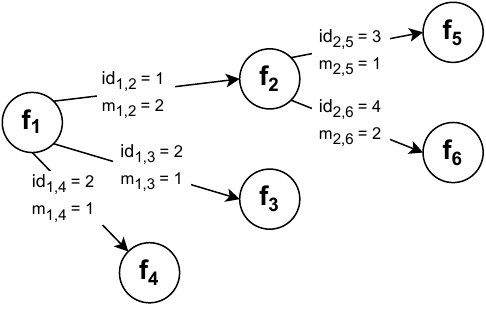}
  \caption{Example of annotated DAG.}
  \label{fig:edges}
\end{figure}

\approach considers the response time $rt_i$ of a function $f_i \in F$ as follows:
\begin{equation}
\label{eq:rt}
rt_i = lrt_i + ert_i
\end{equation}
where $lrt_i$  is the \textit{local} response time spent for executing its code, that is, the set of instructions that implements the function without considering external calls to other functions, and $ert_i$ is the \textit{external} response time, that is, the time spent for invoking other functions.
The external response time depends on the type of dependency between $f_i$ and another function $f_j$. First, the invocation could be either \textit{sequential} or \textit{parallel}. In the former case, the invocation is synchronous, that is, other invocations must wait for its completion. The latter allows it to be executed in parallel with some other invocations. More formally, each edge in the DAG is annotated with an identifier $id_{i,j}$. If two edges $E_{i,j}$ and $E_{i,k}$, which represent the invocations of $f_j$ and $f_k$, respectively, in $f_i$, have different identifiers, the invocations are executed sequentially. If they are annotated with the same value, they are called in parallel.\footnote{\blue{Note that our approach allows a function $f_i$ to invoke another function $f_j$ both in sequential and parallel mode by having multiple, properly annotated, edges between $i$ and $j$ (as in \textit{multigraphs}~\cite{balakrishnan1997graph}). Herein, we did not detail such edge cases to keep our formalization as simple as possible. }} 
Moreover, each edge is also annotated with a \textit{multiplier} $m_{i,j}$, which denotes how many times such invocation is executed within the same function call~\cite{xu2023statefulServ}. \blue{Figure \ref{fig:edges} shows an example of an annotated DAG with $5$ functions. Function $f_1$  sequentially calls function $f_2$ (i.e., $id_{1,2}$ is unique) two times during its execution (i.e., $m_{1,2} = 2$) ---for example, at the beginning and at the end of the computation. Moreover, $f_1$ invokes $f_3$ and $f_4$ in parallel (i.e., $id_{1,3}=id_{1,4}=2$) each only once  (i.e., $m_{1,3} =  m_{1,4} = 1$). Finally, $f_2$ invokes sequentially $f_5$ (once) and $f_6$ (two times) since the edge identifiers are unique.}
The external response time is defined as follows:
\begin{equation}
\label{eq:ert}
  ert_i = \sum_{j \in S} m_{i,j}*rt_j + \sum_{P \in \bar{P}} max(m_{i,j}*rt_j, \, \, \forall j \in P)
\end{equation}
where $S$ is the set of functions that $f_i$ invokes sequentially, $\bar{P}$ is the set of the subsets of functions that $f_i$ invokes in parallel, and $P$ represents each subset (i.e., a group of parallel invocations). In essence, the external response time is equal to the sum of all sequential invocations plus the sum of all the longest response times of each parallel group \blue{taking into account how many times each dependency is called (i.e., $m_{i,j}$)}. 

In \approach, function instances can receive requests from users (i.e., direct invocations)  and/or from other functions (i.e., external invocations). More formally, the total amount of requests $r^w_i$ received by $f_i$ in a set time window $w$ is defined as follows:

\begin{equation}
\label{eq:requestsManagment}
r_i^w = r^w_{users\rightarrow f_i} + \sum_{j \in E_{*,i}} m_{j,i} * r^w_j
\end{equation}
where $r^w_{users\rightarrow f_i}$ is the total amount of direct invocations of $f_i$ in $w$ and $E_{*,i}$ is the set of edges from a source node (i.e., a function $f_j$) to $f_i$. 

\subsection{Control Algorithm}
\label{sec:CoreAllocationNEPTUNE$^+$}

While NEPTUNE's node controllers consider each function independently of the others, \approach adapts the control strategy by considering the dependency DAG. The control algorithm we employ distinguishes between \textit{entrypoint} functions and \textit{externally invoked} functions.  The former can be invoked by users, that is, $r^w_{users\rightarrow f_i} > 0$ for some $w$, whereas the latter are \textit{only} called by other functions in the DAG\footnote{\blue{Note that, in this context, the invocation is \textit{external} with respect to the execution environment of the invoked function (as for the use of ``external'' in \textit{external response time}).}}.  

\approach allows users to define an SLA for each entrypoint function $f_i$. If $f_i$ is only called by users and not by other functions, such an input is mandatory, whereas it is optional if $f_i$ is also invoked by other functions (as a consequence of a direct invocation of another entrypoint).

\approach inherits from NEPTUNE its PI controllers, which compute resource allocations without synchronizing with one another. The main difference between the two approaches is that NEPTUNE monitors and controls the response $rt_i$ of each function $f_i$ \blue{without discriminating between local and external response times}, whereas \approach focuses on the local response time $lrt_i$. The intuition behind this design is that $rt_i$ is affected by the response times of other functions (\blue{external} invocations), which results in the problems described in Section~\ref{sec:netpune:limitations}. In contrast, $lrt_i$ depends solely on the resource allocation of $f_i$ and allows for a more fine-grained and optimized control strategy. 
\begin{figure}[t]
  \centering
\includegraphics[width=0.75\textwidth]{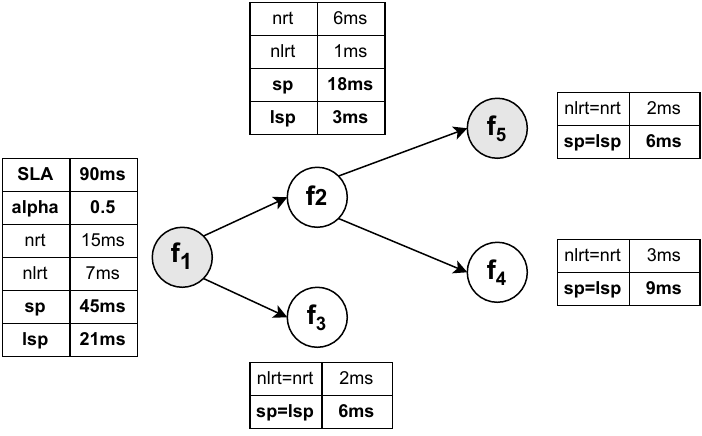}
  \caption{Example of local set point computation.}
  \label{fig:example-st}
\end{figure}
Since \approach exploits $lrt_i$, it cannot simply reuse Equation \ref{eq:setpoint} to define \textit{local set points} for the PI controllers, since $SLA_i$ is defined as an upper bound of the \textit{total} response $rt_i$. Thus, \approach computes, at design time, the local set points $lsp_i$ of each function $f_i$ by considering: i) the user-defined SLAs for entrypoint functions, ii) the dependency among functions, and iii) the \textit{weight} of each function within the DAG. Intuitively, higher weights correspond to higher set points since such functions are considered more complex compared to others.

In particular, the weight of a function $f_i$ is calculated by using the \textit{nominal response time} $nrt_i$ and the \textit{nominal local response time} $nlrt_i$: 
$nrt_i$ and $nlrt_i$ are modeled, respectively, by using the formulas used to calculate $rt_i$ (Equation
\ref{eq:rt}) and $lrt_i$ (Equation \ref{eq:ert}). However, whereas  $rt_i$ and $lrt_i$ are measured at runtime under user-generated workloads,  $nrt_i$ and $nlrt_i$ are measured during a profiling phase while considering the system in a quiescent state (i.e., without saturation or request queue). Each function is profiled with a static core allocation (e.g., $1$ core) and by sending one request at a time, waiting for the previous to finish. To avoid considering cold starts, the measurement starts after a warm-up period. These two metrics are used to understand the complexity of functions and their dependencies in a ``controlled'' state and are used to properly compute set points.

To explain how we calculate local set points, we employ the same application described in Section~\ref{sec:neptune} and  Figure~\ref{fig:example-st} shows the main calculations. Functions $f_1$ and $f_5$ are depicted in light gray and are the two entrypoints of the application. For the sake of simplicity, we consider all the dependencies as sequential and with all multipliers $m_{i,j}$ equal to $1$. \blue{Thus, we avoided reporting the DAG annotation}.

Let $f_i$ be an entrypoint function with a user-defined $SLA_i$, and a set point $sp_i$ defined as in Equation~\ref{eq:setpoint}. In the example, only $f_1$ has a user-defined SLA that is equal to $90ms$. Since $\alpha$ is set to $0.5$, the set point $sp_1$ is equal to $45ms$. The local set point $lsp_i$ is defined as: 

\begin{equation}
\label{eq:localsetpoint}
lsp_i = sp_i * \frac{nlrt_i}{nrt_i}
\end{equation}

Thus, it follows that, in the example, the local set point of function $f_1$ is equal to $21ms$. Moreover, \blue{for each function $f_j$ that is invoked by $f_i$} the set point $sp_j$ is then defined as:

\begin{equation}
\label{eq:setpointdep}
sp_j = \frac{sp_i}{m_{i,j}} * \frac{nrt_j}{nrt_i}
\end{equation}

The setpoint $sp_i$ is used to compute the local set point of $f_i$ and the set points $sp_j$ of all the dependencies. Given that $sp_j$ is intended to consider a single invocation, the calculation is divided by $m_{i,j}$. 

In the example, the dependencies of $f_1$ are $f_2$ and $f_3$. This means that, ideally, the sum of the local response time of $f_1$ and the \blue{(total)} response times of $f_2$ and $f_3$ should be equal to set point $sp_1$. Since $f_3$ has no dependencies, \blue{its nominal response time $nrt_3$ is equal to its nominal local response time $nlrt_3$}. Intuitively, since $nrt_3$ is around $3$ times lower than $nlrt_1$ the set point of $f_3$ ($6ms$) is roughly $3$ times lower than the local set point of $f_1$ ($21ms$). Instead, $f_2$ depends on $f_4$ and $f_5$, and its set point $sp_2$ is set to $18ms$ with a nominal response time $nrt_2$ equal to $6ms$. This means, in turn, that the sum of the local response time of $f_2$ and the response times of $f_4$ and $f_5$ should be kept equal to $18ms$. 

Recursively, the local set points of each dependency are calculated by using Equation \ref{eq:localsetpoint}. For example, the set point $sp_2$ is used to calculate the local set point of $f_2$ ($3ms$) along with the set points of $f_4$ ($6ms$) and $f_5$ ($9ms$). Note that the case of a function without dependencies is the base case of the recursive procedure, and \blue{its set point is equal to its local set point}, such as for $f_3$, $f_4$, and $f_5$. 
Finally, to further optimize the resource allocation, the set points of parallel dependencies (i.e., edges with the same source node and identifiers) are calculated as in Equation \ref{eq:setpointdep}. However, after the calculation, the set point of each dependency is set to be equal to the maximum of the parallel group. This means that even though siblings could complete execution faster, they are slowed down with higher set points to match the slowest function of the parallel group. This strategy does not affect the overall response time of the application and allows for saving resources.

\paragraph{Proportional-Integral Control.}
After the previous steps, each function is provided with a local set point $lsp_i$. As in NEPTUNE, each function $f_i$ is equipped with a PI controller. In \approach, the controller for a function $f_i$ monitors only the local response time $lrt_i$ and allocates cores to meet $lsp_i$. Without any synchronization and thanks to the computations above, if all the controllers are able to meet their local set points, user-defined SLAs are fulfilled. The control algorithm we used, adapted from NEPTUNE, is reported in Algorithm \ref{alg:Algorithm3}.

\begin{algorithm}[t]
\caption{Core allocation}
\label{alg:Algorithm3}
\begin{algorithmic}[1]
    \State $lrt_i \gets getLocalResponseTime(f_i)$
    \State $err \gets lsp_{i}^{-1} - lrt_{i}^{-1}$
	\State $P \gets gain_{P} * err$
	\State $I \gets I + gain_{I} * err$
    \State $cores \gets P + I$
	\State $cores \gets min( cores_{MAX}, max(cores_{MIN}, cores)) $
    \State $allocateCores(f_i, cores)$
\end{algorithmic}
\end{algorithm}
The procedure is invoked at every control period for each function instance $f_i$.
The local response time (obtained at line $1$) and the local set point are used to compute the error $err$. The higher the error is, the higher the mismatch between the local set point and the actual measured local response time (line $2$). The proportional contribution ($P$) is equal to the proportional gain $gain_P$ multiplied by $err$ (line 3). The integral contribution ($I$) is the sum of the previous actions and the error times the integral gain $gain_I$ (line $4$). Both $gain_P$ and $gain_I$ are tuning parameters of the controller and can be set using different well-known heuristics \cite{borase2021review}.
The core allocation is computed as the sum of $P$ and $I$ (line $5$) properly scaled according to the minimum ($cores_{MIN}$) and maximum ($cores_{MAX}$) allowed amount of cores (line $6$). Finally, the allocation is enacted at line $7$.

\section{Evaluation}
\label{sec:evaluation}
Our evaluation was aimed to compare \approach against NEPTUNE in the absence of bottlenecks and when bottlenecks occur.

We ran all the experiments on a MacBook Pro equipped with $4$ cores and $16$GB of RAM  running macOS Ventura (version $13.2.1$).
To test the two systems, we relied on an existing simulator called RAS (Resource Allocation Simulator), created by the same authors of NEPTUNE~\cite{baresi2020simulationCloudApps}. The RAS simulator was originally used to evaluate the performance of the control algorithms of NEPTUNE against industrial approaches. We extended the simulator~\footnote{Source code available at \url{https://doi.org/10.5281/zenodo.8174489}} in two ways: i) we adapted the code to support function dependencies, and ii) we implemented the novel theoretical model and control algorithm. 

The experiments used three different applications. The first two are benchmarks widely used in the literature~\cite{Hossen2022autoscal}, namely \textit{hotel reservation}\footnote{https://github.com/vhive-serverless/vSwarm/tree/main/benchmarks/hotel-app} and \textit{sockshop}\footnote{https://github.com/microservices-demo/microservices-demo}). The first is a serverless application that mimics a hotel reservation website, whereas the second is an online e-commerce application that exploits a microservice architecture that was converted to serverless functions by the authors of NEPTUNE~\cite{baresi2022neptune}. \textit{Hotel reservation} includes four functions ($2$ entrypoints), and it is characterized by a DAG with an average out-degree of $3$ edges, an average in-degree of $1$ edge, and all the dependencies have type \textit{sequential}. \textit{Sockshop} includes $7$ functions ($5$ entrypoints) with an average $out-degree$ of $6$ edges, an average $in-degree$ of $1$ edge, and one third of the dependencies have type \textit{sequencial} while the remaining two thirds have type \textit{parallel}.  We also created a more complex scenario, application \textit{complex}, by synthesizing a DAG of $25$ functions ($6$ entrypoints), with an average out-degree of $2$ edges, an average in-degree of $1$ edge, and roughly balanced sequential and parallel dependencies.

We repeated each experiment $10$ times. In each test, we simulated executions of 20 minutes each, and we collected,  for each function, the average ($\mu$) and the standard deviation ($\sigma$) of three metrics: response times ($RT$) in milliseconds, core allocations ($C$) in millicores, and percentage of SLA violations ($V$).

The tests employed workloads similar to the ones used to evaluate NEPTUNE in~\cite{baresi2020simulationCloudApps,baresi2022neptune}. In particular, each entrypoint function was stimulated with either a \textit{ramp} or a step \textit{workload}. We employed ramps that start from $10$ requests and added one request every second up to $100$ (as in \cite{baresi2022neptune}) and randomly generated steps that vary the workload every $50$ seconds in a range between $20$ and $120$ requests. We also simulated bottlenecks by changing the random step to a number of requests that ranges between $800$ and $6000$.

For NEPTUNE we set an SLA for each function, whereas for 
\approach we only set them for entrypoints, since our approach is able to automatically calculate the set points for all the other functions. For \textit{sock-shop}, we employed the same SLAs reported in the original NEPTUNE paper~\cite{baresi2022neptune}. For \textit{hotel reservation} and \textit{complex application }, we set the SLAs to double their nominal response times. The nominal response times of \textit{hotel reservation} were obtained by profiling each function, while for \textit{complex application} we generated them randomly.

We configured both NEPTUNE and \approach the same way. We employed a value for $\alpha$ equal to $0.5$ for each function with SLA as in ~\cite{baresi2022neptune}. We derived the values of $gain_P$ and $gain_I$ through manual tuning (again, as in~\cite{baresi2022neptune}).

To be sure that the simulator was aligned with realistic results and that our modifications did not affect its accuracy, we executed a preliminary experiment. We simulated the same tests on NEPTUNE run  in~\cite{baresi2022neptune} with application \textit{sockshop} (i.e., same workload and configuration). We collected the results and compared them with those reported in the paper. We observed that on average the differences were minimal: $0.3\%$ for response times and $4.3\%$ for core allocation.

\subsection{Performance without bottlenecks}

Table \ref{tab:results} shows the results obtained by NEPTUNE (\textit{N}) and \approach (\textit{N+}). For the first two applications, the table lists the tested functions along with their SLAs. For application  \textit{complex}, it only shows averages due to lack of space. Functions marked with a $*$  are entrypoints. Row \textit{overall} reports the averages over all functions. 

If we focus on the part \textit{without bottlenecks}, one can observe that \approach consistently outperforms NEPTUNE in many cases.

For example, if we consider function \textit{order} of \textit{sockshop} application, \approach yields \blue{a significantly more efficient resource allocation compared to NEPTUNE ($788$ millicores allocated by \approach against $1133$ millicores allocated by NEPTUNE)}. \blue{The response time  of \approach ($291.3$ ms) is closer to the $300$ ms set point ($\alpha*SLA$ with $\alpha=0.5$ and $SLA=600ms$)  compared to the result obtained by NEPTUNE ($211.2$ ms). This means that \approach does not need to over-provision CPU cores to meet the user-defined SLA and only allocates needed resources.}
Overall, with benchmark \textit{sockshop}, \approach demonstrates a more efficient performance by reducing required millicores from $510$ to $388$, marking a 24\% reduction, while only having a small increase in average response time ($85.6$ vs. $63.1$ ms) and no SLA violations. \blue{Note that, faster response times can be also obtained by \approach by simply lowering the set points.}

\begin{table}[t]
\footnotesize
\centering
\caption{Results without and with bottlenecks.}
\setlength{\tabcolsep}{1.1pt}
\label{tab:results}
\vspace{0.2cm}
\begin{tabular}{rcr|cccccc|cccccc}
\multicolumn{3}{c}{} & \multicolumn{6}{c}{\textbf{without bottlenecks}} & \multicolumn{6}{c}{\textbf{with bottlenecks}} \\  

\multicolumn{1}{c}{\boldmath$f$} & \multicolumn{1}{c}{\textbf{SLA}} & & \multicolumn{2}{c}{\textbf{RT}} & \multicolumn{2}{c}{\textbf{V}} & \multicolumn{2}{c}{\textbf{C}} & \multicolumn{2}{c}{\textbf{RT}} & \multicolumn{2}{c}{\textbf{V}} & \multicolumn{2}{c}{\textbf{C}} \\  

 & & &  \textit{N} & \textit{N+} &  \textit{N} & \textit{N+} & \textit{N }& \textit{N+} &  \textit{N} & \textit{N+} &  \textit{N} & \textit{N+} & \textit{N }& \textit{N+} \\  \hline
& & & \multicolumn{12}{c}{\textbf{\textit{hotel reservation}}} \\
search* & 118 & $\mu$ & 56.6 & 62.6 & 0 & 0 & 327 & 282 & 65 & 78.2 & 0 & 0 & 2719 & 282 \\
 &  & $\sigma$ & 0 & 0.1 & 0 & 0 & 0.2 & 0 & 0 & 0 & 0 & 0 & 4 & 0 \\
profile* & 36 & $\mu$ & 17.3 & 17.2 & 0 & 0 & 343 & 346 & 33 & 33 & 46.6 & 46.5 & 2285 & 2290 \\
 &  & $\sigma$ & 0 & 0 & 0 & 0 & 0.2 & 0.2 & 0 & 0 & 0.1 & 0.1 & 0.7 & 0.7 \\
geo & 27 & $\mu$ & 12.9 & 12.8 & 0 & 0 & 243 & 245 & 12.9 & 12.8 & 0 & 0 & 243 & 245 \\
 &  & $\sigma$ & 0 & 0 & 0 & 0 & 0.1 & 0.1 & 0 & 0 & 0 & 0 & 0.1 & 0.1 \\
rate & 34 & $\mu$ & 16.3 & 16.3 & 0 & 0 & 295 & 295 & 16.3 & 16.3 & 0 & 0 & 295 & 295 \\
 &  & $\sigma$ & 0 & 0 & 0 & 0 & 0.2 & 0.2 & 0 
 & 0 & 0 & 0 & 0.1 & 0.1 \\

\multicolumn{2}{c}{\textbf{overall}}  & \textbf{$\mu$} & \textbf{25.8} & \textbf{27.2} &\textbf{0} & \textbf{0} & \textbf{302} & \textbf{292} & \textbf{31.8} & \textbf{35} & \textbf{11.7} & \textbf{11.6} & \textbf{1386} & \textbf{778}\\
 
 & & & \multicolumn{12}{c}{\textbf{\textit{sockshop}}} \\
orders* & 600 & $\mu$ & 211.2 & 291.3 & 0 & 0 & 1133 & 788 & 317.3 & 384.7 & 0 & 0 & 2093 & 788.3 \\
 &  & $\sigma$ & 0.1 & 0 & 0 & 0 & 0 & 0.6 & 0.1 & 0.1 & 0 & 0 & 5.2 & 0.5 \\
catalogue* & 200 & $\mu$ & 50.6 & 72.4 & 0 & 0 & 126 & 88 & 50.6 & 72.4 & 0 & 0 & 126 & 88 \\
 &  & $\sigma$ & 0 & 0 & 0 & 0 & 0 & 0.1 & 0.1 & 0 & 0 & 0 & 0 & 0.1 \\
shipping & 50 & $\mu$ & 15.4 & 20.6 & 0 & 0 & 414 & 312 & 15.4 & 20.6 & 0 & 0 & 414 & 312 \\
 &  & $\sigma$ & 0 & 0 & 0 & 0 & 0 & 0.1 & 0 & 0 & 0 & 0 & 0 & 0.2 \\
users* & 50 & $\mu$ & 24.1 & 29.7 & 0 & 0 & 154 & 127 & 24.1 & 29.7 & 0 & 0 & 154 & 127 \\
 &  & $\sigma$ & 0 & 0 & 0 & 0 & 0.1 & 0.1 & 0 & 0 & 0 & 0 & 0.1 & 0.1 \\
payment* & 50 & $\mu$ & 13.7 & 14.4 & 0 & 0 & 605 & 578 & 13.7 & 14.4 & 0 & 0 & 605 & 578 \\
 &  & $\sigma$ & 0 & 0 & 0 & 0 & 0 & 0.6 & 0 & 0 & 0 & 0 & 0 & 0.5 \\
cart-utils & 200 & $\mu$ & 58.8 & 79 & 0 & 0 & 511 & 372 & 58.9 & 79 & 0 & 0 & 511 & 372 \\
 &  & $\sigma$ & 0 & 0 & 0 & 0 & 0 & 0.4 & 0 & 0 & 0 & 0 & 0 & 0.3 \\
cart-del* & 200 & $\mu$ & 67.8 & 92 & 0 & 0 & 628 & 450 & 185.8 & 185.4 & 45.9 & 45.9 & 1527 & 1530 \\
 &  & $\sigma$ & 0 & 0 & 0 & 0 & 0 & 0.3 & 0.1 & 0.1 & 0 & 0 & 0.2 & 0.2 \\
 \multicolumn{2}{c}{\textbf{overall}} &  \textbf{$\mu$} & \textbf{63.1} & \textbf{85.6} &\textbf{0} & \textbf{0} & \textbf{510} & \textbf{388} & \textbf{95.1} & \textbf{112.3} & \textbf{6.6} & \textbf{6.6} & \textbf{776} & \textbf{542}\\
 & & & \multicolumn{12}{c}{\textbf{\textit{complex}}} \\
 \multicolumn{2}{c}{\textbf{overall}}  & \textbf{$\mu$} & \textbf{240.0} & \textbf{263.7} &\textbf{0} & \textbf{0} & \textbf{4627} & \textbf{3013} & \textbf{313.2} & \textbf{357.5} & \textbf{11.5} & \textbf{15.3} & \textbf{6530} & \textbf{3760}\\
 \hline
\end{tabular}
\end{table}

The trend is similar with more complex applications (i.e., \textit{complex}): \approach yields a more efficient allocation ($3013$ vs. $4627$ millicores), with a 27\% improvement, \blue{no SLA violations, and comparable to NEPTUNE response times.}

\blue{Conversely, the two approaches provide similar performance with benchmark \textit{hotel reservation} except for function \textit{search} where \approach is slightly more efficient in terms of core allocation. This can be attributed to the application's simple DAG and its limited amount of dependencies. This result demonstrates that \approach does not introduce any performance degradation in scenarios where dependencies are not a critical factor.} 

\subsection{Performance with bottlenecks}

\begin{figure}[t]
    \centering
    \begin{subfigure}[t]{0.495\textwidth}
\includegraphics[width=\textwidth]{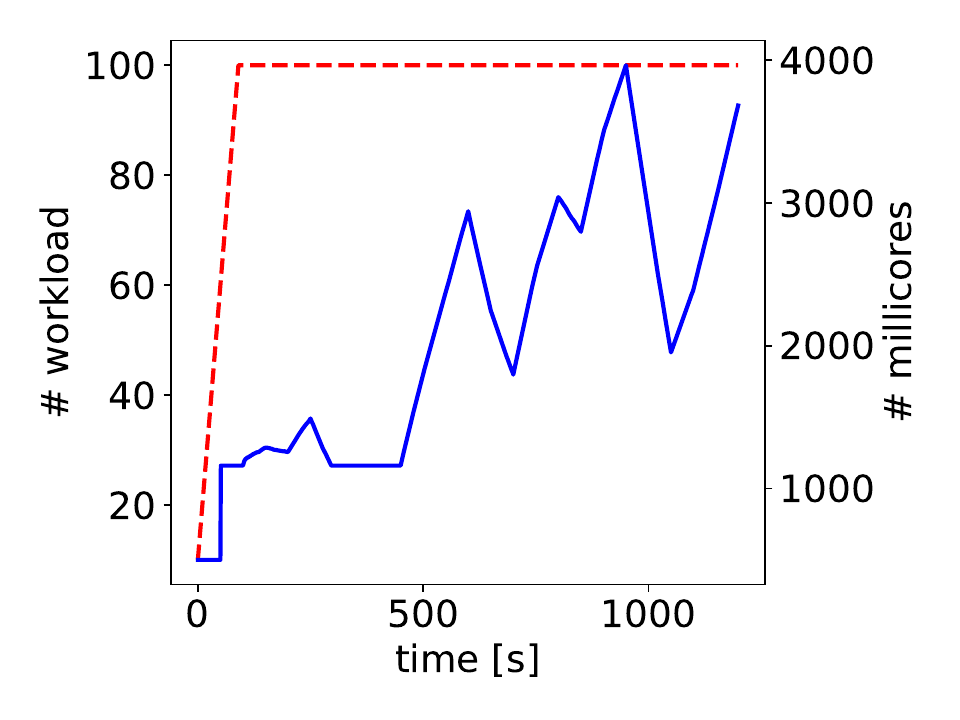}
        \caption{NEPTUNE}
        \label{fig:image1}
    \end{subfigure}
    \hfill
    \begin{subfigure}[t]{0.495\textwidth}
\includegraphics[width=\textwidth]{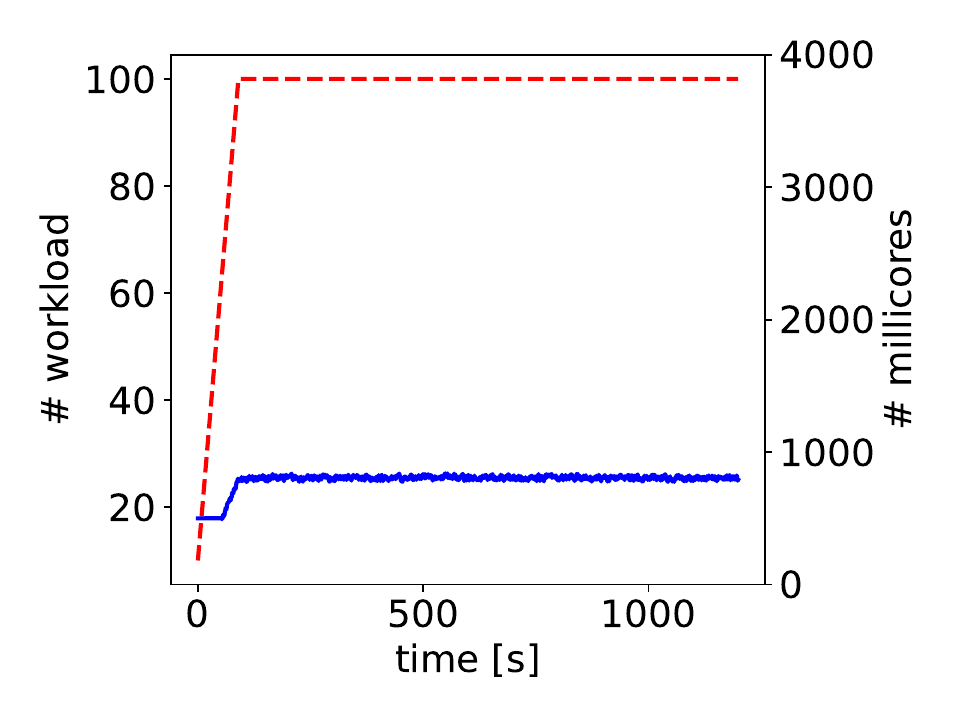}
        \caption{\approach}
        \label{fig:image2}
    \end{subfigure}
    \hfill
    \caption{Results for function \textit{order} (\textit{sockshop}). }
    \label{fig:cores}
\end{figure}

Table \ref{tab:results} also shows the results obtained when managing bottlenecks (created as explained above). As for application \textit{hotel reservation}, we raised the number of requests for function \textit{profile}, leading to a significant amount (around 47\%) of SLA violations obtained by both NEPTUNE and \approach. Such a bottleneck inevitably raises the response time of \textit{search}, which directly depends on \textit{profile}. Since \approach only considers local response times, our solution is able to properly manage this function by only allocating $282$ millicores on average, while NEPTUNE raises the average core allocation to $2719$ millicores, that is, some $90\%$ higher than \approach. The response times are comparable: $65 ms$ with NEPTUNE and $78.2 ms$ \approach. The other functions, not affected by the bottleneck, showed performance similar to the ones observed in the experiments described in the previous section. Overall with this application, \approach obtained a slightly higher average response time  ($35 ms$ vs. $31.8 ms$) and a $44\%$ lower average core allocation ($778$  vs. $1386$ millicores).

We observed similar results also with \textit{sockshop}. In this case, we created a bottleneck in function \textit{cart-del} as demonstrated by the high number of violations obtained by the two approaches. The results reported for function \textit{order}, which depends on \textit{cart-del}, clearly show the benefits of \approach. While \approach results in a higher response time ($384.7$ vs. $317.3$), it also significantly reduces the number of cores used, allocating only $788.3$ millicores against $2093$ (62\% improvement). This is clearly shown in Figure~\ref{fig:cores}, where NEPTUNE's controller for function \textit{order} is unstable due to the bottleneck in \textit{cart-del} and reaches a peak of some $4000$ allocated millicores. In contrast, \approach keeps its allocation roughly stable at around $800$ millicores after the initial ramp.  Overall, as for \textit{sockshop}, \approach obtained a core allocation that is almost 30\% better (lower) on average than NEPTUNE with comparable response times (equal SLA violations).

\blue{Application \textit{complex} suggests that the more complex an application becomes, the more efficient \approach is: $3760$ vs. $6530$ millicores, with a 42\% improvement at the cost of only 4.8\% more SLA violations.}

\blue{By taking into account function dependencies, \approach efficiently allocates resources across diverse benchmarks and scenarios. Conversely, NEPTUNE, without dependency awareness, tends to over-provision resources. This results in faster, yet less optimized, response times that only rarely lead to fewer SLA violations. NEPTUNE's behavior is partly due to its inability to maintain set points, resulting in an over-speeding that is not beneficial in most of the cases. Instead, \approach provides more precise control and offers a more convenient trade-off between resource efficiency and response times. If faster response times are required, \approach users can simply define stricter SLAs or lower the value of $\alpha$ to obtain a more responsive system.}


\section{Related Work}
\label{sec:related}
The problem of managing microservices or serverless functions deployed on edge infrastructures has been already studied in the literature~\cite{wang2021lass,baresi2022neptune}. Such approaches tackle component placement, routing, and resource management, but only a few of them take function dependency into account~\cite{he2022online,mahgoub2021sonic,xu2023statefulServ}.

He et al.~\cite{he2022online} introduce a novel approach for deploying microservices to edge servers by taking into account their intricate dependencies with the goal of optimizing response time. They do not consider resource allocation but only component placement. Therefore, the amount of CPU cores to obtain a certain response time is not optimized. In contrast, \approach considers the trade-off between allocated resources and response time. They also do not consider parallel and multiple invocations of the same components.

Xu et al.~\cite{xu2023statefulServ} propose a solution for optimizing the placement at the edge of serverless functions with dependencies. They also take into account stateful computation and network delays. Similarly, Ashraf et al.~\cite{mahgoub2021sonic} propose SONIC, a solution that aims to optimize the performance and operation cost of serverless applications by deciding the best function placement for exchanging data. Applications are abstracted as DAGs as in \approach. These approaches select the best path for exchanging data by considering data size, function dependencies, and network state. They are complementary to \approach since they exploit dependencies for optimizing data exchange and do not consider resource allocation. 

Moving to runtime resource management~\cite{bhasi2021Kraken,daw2020xanadu,wang2021lass,zuk2022reduceRtCompositeFaaS}, these studies either disregard function dependencies entirely, as in the case of~\cite{wang2021lass}, or they utilize a probabilistic approach to pre-allocate functions, such as~\cite{bhasi2021Kraken,daw2020xanadu,zuk2022reduceRtCompositeFaaS}. 
For instance, Daw et al.~\cite{daw2020xanadu} introduce Xanadu, which uses a directed acyclic graph of dependencies and a probabilistic model to identify the most likely execution paths. To reduce the overhead from the cascading cold-start of functions, it pre-allocates resources (i.e., containers) for the most probable path in response to each function call. 
Similarly, Kraken~\cite{bhasi2021Kraken} estimates the request volume for each function and uses these estimates to deploy the necessary containers. By batching multiple requests, Kraken utilizes fewer resources than Xanadu: under moderate to heavy load, Xanadu deploys nearly 32\% more containers than Kraken~\cite{bhasi2021Kraken}. However, both Xanadu and Kraken's probabilistic approaches can lead to resource over- or under-provisioning if the estimated probability distribution does not reflect the actual workload fluctuations over time. Furthermore, neither solution considers resource allocation for functions with dependencies in the event of a bottleneck as \approach does.

Conversely, Wang et al.~\cite{wang2021lass} present LaSS, a platform for managing the latency of serverless computations. LaSS uses a dynamic resource allocation strategy based on workload variations and a queuing model. A weighted fair-share resource allocation strategy is employed to prevent overload and maintain the desired response time. While this work makes a significant contribution by mitigating SLA violations and over-allocation of resources, the authors do not consider function dependencies, which could lead to inefficient allocations. Once more, \approach uses control theory to only allocate the necessary resources to functions, based on the number of requests and defined SLA, and allows for a more efficient resource usage.

\section{Conclusions}
\label{sec:conclusions}
This paper presents \approach, a dependency-aware resource allocation solution for serverless functions deployed on edge infrastructures. We extended  NEPTUNE by developing a new theoretical model and control algorithm that exploit dependencies to efficiently allocate CPU cores to serverless functions.  The evaluation shows that \approach outperforms the original framework up to 42\% in terms of resource allocation. In the future, we will improve our solution by also considering the placement of dependency-aware functions.

\bibliographystyle{plain}
\bibliography{main}
\end{document}